\documentclass[universe,review,accept,moreauthors,12pt,a4paper]{mdpi} 
\lastpage{x}
\doinum{10.3390/------}
\pubvolume{xx}
\pubyear{}
\externaleditor{Academic Editor: Kazuharu Bamba}
\history{Received: 15 July 2015 / Accepted: 22 July 2015 / Published: xx}

\usepackage{graphicx}
\usepackage{booktabs}
\usepackage{multirow}
\usepackage{amssymb}
\usepackage{lscape}
\usepackage{longtable}
\usepackage{rotating}
\usepackage{colordvi}
\usepackage{color}
\usepackage[english]{babel}
\usepackage{epsfig}
\usepackage{subfig}
\usepackage{textcomp}
\usepackage{natbib}
\usepackage{soul}
\usepackage{enumitem}

\Title{Constraining $f(R)$ Gravity by the Large-Scale Structure}

\Author{Ivan de Martino $^{1,}$*, Mariafelicia De Laurentis $^{2,3}$ and Salvatore Capozziello $^{3,4,5}$}

\address{%
$^{1}$ F\'{i}sica Te\'orica, Universidad de Salamanca, 37008 Salamanca, Spain\\
$^{2}$ Tomsk State Pedagogical University, ul. Kievskaya, 60, 634061 Tomsk, Russia; \linebreak E-Mail: mfdelaurentis@tspu.edu.ru\\
$^{3}$ Istituto Nazionale di Fisica Nucleare (INFN) Sez. di Napoli, Complesso Universitario di Monte Sant'Angelo, 
Edificio G, via Cinthia, I-80126 Napoli, Italy; \linebreak E-Mail: capozziello@na.infn.it \\
$^{4}$ Dipartimento di Fisica, Universit\'a di Napoli ``Federico II'', Complesso Universitario di Monte Sant'Angelo, Edificio~G, 
via Cinthia, I-80126 Napoli, Italy\\
$^{5}$ Gran Sasso Science Institute (INFN), via F. Crispi 7, I-67100 L' Aquila, Italy \vspace{-12pt}}

\corres{E-Mail: ivan.demartino1983@gmail.com; Tel.: 0034 923 294 437 }

\abstract{
Over the past few decades, general relativity and the concordance $\Lambda$CDM model have been successfully tested 
using several different astrophysical and cosmological probes based on large datasets ({precision cosmology}). Despite their successes, some shortcomings emerge due to the fact that general relativity should be revised at infrared and ultraviolet limits and to the fact that the fundamental nature of dark matter and dark energy is still a puzzle to be solved. In this perspective, $f(R)$ gravity has been extensively investigated, being the most straightforward way to modify general relativity and to overcame some of the above shortcomings. In this paper, we review various aspects of $f(R)$ gravity at extragalactic and cosmological levels. In particular, we consider a cluster of galaxies, cosmological perturbations and N-body simulations, focusing on those models that satisfy both cosmological and local gravity constraints. The perspective is that some classes of $f(R)$ models can be consistently constrained by the large-scale structure.}


\begin{document}

\section{Introduction}\label{sec:intro}

The {concordance} $\Lambda$CDM cosmological model is based on Einstein's
general relativity (GR), the~standard model of particles with the inclusion of 
two new ingredients, which are the cosmological constant $\Lambda$ and dark matter. 
It provides a theoretical framework to describe the
formation and the evolution of the structures in the Universe.
It has been successfully tested using many different
datasets, such as the Supernovae Type Ia (SNeIa)
\cite{Perlmutter1997,Riess2004, Astier2006, Suzuki2012}, the
matter power spectrum from the two-degree field (2dF) survey of galaxies \cite{Percival2001, Pope2004}
and from the Sloan Digital Sky Survey (SDSS) \cite{Tegmark2004}, 
the temperature fluctuations of the Cosmic Microwave Background (CMB) radiation 
\cite{Hinshaw2013, Planck2013_15, Planck2013_20, Planck2013_23, Planck2015_1, Planck2015_13,
Planck2015_14, Planck2015_17, Planck2015_18, Planck2015_20, Planck2015_21},
the baryonic acoustic oscillations (BAO) \cite{Blake2011},
and so on. 
Despite its successes, the following questions are still need to be answered: Is GR sufficient to explain all 
of the gravitational phenomena from collapsed objects
to the evolution of the Universe and the formation of the structures? Does the theory
need to be changed, or at least modified? There exist two different approaches to 
solve the debate: preserve the successes of GR by incorporating
new particles and/or scalar fields not yet observed and improve the accuracy of the
data to further verify the model; or modify the theory of gravity to make it compatible 
with quantum mechanics and cosmological observations without introducing
additional particles and fields. Independently of the preferred approach, 
having alternatives demands testing the model more deeply.
In the {concordance} $\Lambda$CDM model, the most important energy density component is 
the cosmological constant $\Lambda$ (or more in general, the dark energy (DE)), 
required to explain the accelerated expansion of the Universe, with
$\Omega_\Lambda=0.691\pm0.006$ \cite{Planck2015_13} in units of the critical density.
The second in importance is dark matter (DM), with $\Omega_{DM}= 0.259\pm0.005$ \cite{Planck2015_13},
needed to explain the emergence of the large-scale structure (LSS). 
However, the lack of a full comprehension
of the fundamental nature of those two components, whether particles or scalar fields, is completely unknown and
has demanded to further test the GR to understand if it is the effective theory of gravity.
 Additionally, the need for requiring two unknown components to fully explain the astrophysical and cosmological 
observations has been interpreted as a breakdown of the theory at those scales \cite{CapDeLa2011}. 

As an alternative to introducing extra matter components in the stress-energy tensor, it is possible to change the geometrical description of the gravitational interaction, adding in the Hilbert--Einstein Lagrangian, higher-order curvature invariants, such as $R^{2}$, $R_{\mu\nu} R^{\mu\nu}$, 
$R^{\mu\nu\alpha\beta}R_{\mu\nu\alpha\beta}$, $R \, \Box R$ or $R
\, \Box^{k}R$), and minimally or non-minimally coupled terms between scalar fields
and geometry (such as $\phi^{2}R$)~\cite{CapDeLa2011, OdintsovPR,Mauro,faraoni,olmo,lobo,defelice,sotiriou,scolar}. 
Extended theories of gravity (ETGs)
can be classified as: scalar-tensor theories, when the geometry is non-
minimally coupled to some scalar field; and higher order theories, when the action
contains derivatives of the metric components of order higher than the second.
ETGs differ from GR in many concrete aspects and, therefore, are testable with current 
and forthcoming experiments.
The most straightforward way to extend the theory of gravitation is 
replacing the Einstein--Hilbert Lagrangian, which
is linear in the Ricci scalar, $R$, with a more general function of the curvature $f(R)$.
Since the~exact functional form of the $f(R)$-Lagrangian is unknown, 
theoretical predictions should be matched to all available data (at all scales) 
to accommodate the observations.
$f(R)$ models have been used to test: stellar formation and evolution \cite{Stabile2012, CapDeLaDeMa2012,
Arbuzova2014, Farinelli2014, Astashenok2015a, Astashenok2015b};
emission of gravitational waves and constraints on the massive graviton modes 
\cite{DeMartino2013, DeMartino2015, quadrupolo, DeLaurentisVIRGO, greci, af+13, barrow, clifton, antoniadis,Berry2011};
the flat rotation curves of spiral galaxies \cite{Cardone2011}; 
the velocity dispersion of elliptical galaxies \cite{Napolitano2012}; 
to describe a cluster of galaxies 
\cite{Capozziello2009, Schimdt2009, Ferraro2011, DeMartino2014, Terukina2012,Terukina2014, Wilcox2015}; 
the structure formation and the evolution of the Universe 
\cite{Planck2015_14, Euclid2013, Jennings2012,Zhao2011,Puchwein2013,Arnold2015, EFTCAMB1,Marchini2013,giappo,kazuserge,Carloni2009,Carloni2005,Abdelwahab2008}. 
Although all tests provide clear indications that the $f(R)$-gravity models could
overcame the shortcomings of the $\Lambda$CDM model, a final and self-consistent 
framework has not been reached yet. 

In addition to the simple $f(R)$ model, it is possible to construct much more complex Lagrangians.
As an example, teleparallel gravity considers Lagrangians that are a function of torsion scalar $f(T)$ 
\cite{pereira}. In such models, the Ricci scalar is assumed to vanish, and its role is played by torsion. 
Furthermore, the particularity of these $f(T)$ models is that they include torsion as the source of DE 
without considering a cosmological constant, and they give rise to second order field equations that are 
easier to solve than the fourth order ones from $f(R)$ gravity 
\cite{bambasan,barrowsot, pereira, storny,storny1,storny2,storny3,storny4,storny5,storny6,bamba,basilakosNOI}. 
These models have been widely investigated from fundamental to cosmological scales 
\cite{14,15,16, bambagw,28B,27B,barrowsot,34B}. Recently, many efforts been devoted to investigating the 
$f(R, T)$ models, where the gravitational Lagrangian consists of an arbitrary function of the Ricci scalar 
$R$ and the torsion scalar $T$ \cite{MIORT,34RT,36RT,37RT}. {Another class of interesting models that have been 
introduced} are, for example, $f({\cal G})$, where ${\cal G}$ is Gauss--Bonnet topological invariant or 
combinations of these with the Ricci scalar as the $f(R, {\cal G})$ \cite{18G,19G,20G,21G,22G,23G,24G,25G}. 
Using these models, it is possible to construct viable cosmological models, preventing ghost contributions 
and contributing to the regularization of the gravitational action. 

In this review, we will focus only on $f(R)$ models. We summarize the main achievements, pointing out the 
difficulties and the future perspectives in probing the $f(R)$-gravity using LSS datasets. In Section~\ref{sec.2}, we report the general formalism and the main features of $f(R)$-gravity models, and also, we 
focus on two models/approaches 
that have led to several results in the last few years. In Section \ref{sec.3}, we describe two~approaches to probe $f(R)$-gravity using a cluster of galaxies. In Section \ref{sec.4}, we review the most
important results obtained using N-body simulations to quantitatively describe the physical effect
of $f(R)$-gravity models on the structure formation. In Section \ref{sec.5}, we summarize the main approaches to
test alternative models using the expansion history of the Universe. In Section \ref{sec.6}, we consider the constraints
obtained using the CMB power spectrum. Finally in Section \ref{sec.7}, we give 
the main conclusions and the future perspectives for this field.

\section{{\em f(R)} Gravity}
\label{sec.2}

In metric formalism, the simplest four-dimensional action in $f(R)$ gravity is:
\begin{equation}\label{action}
{\cal A}=\frac{c^4}{16\pi G}\int d^4x \sqrt{-g}\left[f(R) + {\cal L}_{{\rm
matter}}\right]\,
\end{equation}
where $g$ is the determinant of the metric $g_{\mu\nu}$, and ${\cal L}_{\rm
matter}$ is the standard perfect fluid matter Lagrangian. Varying the action Equation~\eqref{action} 
with respect to the metric tensor, we obtain the field equations and its~trace:
\begin{equation}
f_{,R} R_{\mu\nu}-\frac{1}{2}f(R) g_{\mu\nu}-\left[\nabla_{\mu} \nabla_{\nu} - 
g_{\mu\nu} \Box\right]f_{,R}=\frac{8\pi G}{c^{4}} T_{\mu \nu }
\label{field_eq}
\end{equation}
\begin{equation}
f_{,R}R-2 f(R)+3 \Box f_{,R}=\frac{8\pi G}{ c^{4}}T\,
\label{trace_field_eq}
\end{equation}

Here, $f_{,R}=df/dR$, $\Box=g^{\mu\nu }\nabla_\mu \nabla_\nu$ is the d'Alembert operator, 
and $\displaystyle{T_{\mu\nu}= \frac{-2}{\sqrt{-g}}\frac{\delta\left(\sqrt{-g}{\cal L}_m\right)}{\delta g^{\mu\nu}}}$ 
is the energy momentum tensor of the matter.
We can recast the Equation~\eqref{field_eq} in Einstein form as follows:

\begin{eqnarray}\label{EINS}
G_{\mu\nu}&=&\frac{1}{f_{,R}}\left\{\frac{1}{2}g_{\mu\nu}\left[f(R)-Rf_{,R}\right]
+(f_{,R})_{;\mu\nu} -g_{\mu\nu}\Box f_{,R}\right\}+ \frac{\kappa T^{(m)}_{\mu\nu}}{f_{,R}}=\nonumber\\
&&=T^{(curv)}_{\mu\nu}+\frac{T^{(m)}_{\mu\nu}}{f'(R)}\,
\end{eqnarray} 
where we can reinterpret the first term on the right side of the equation as an extra stress-energy 
tensor contribution $T^{(curv)}_{\mu\nu}$. $f(R)$ gravity is sufficiently general
to include all of the basic features of ETGs, but at the same time, it can be easily connected to the observations. 
Let us consider now two classes of $f(R)$ models, which are designed to satisfy cosmological and Solar System constraints.

\subsection{Chameleon Models}\label{sub:chama}

One way to modify gravity is to introduce a scalar field ($\phi$), which is coupled to the matter components 
of the Universe. The range and the length of interactions mediated by the scalar field depend on the density 
of the environment. In high-density configurations, 
the effective mass of the scalar field is large enough to suppress it and to accommodate
all existing constraints from Solar System tests of gravity. Since the mass depends
on the local matter density, the cosmological evolution of the scalar field could be an 
explanation for the acceleration of the Universe. Many environment-dependent screenings have been proposed, such as the chameleon \cite{Khoury2004}, the dilaton \cite{Gasperini2002}, the symmetron \cite{Hinterbichler} or the \mbox{Vainshtein~\cite{Vainshtein, Deffayet}} mechanisms. 
One of the most tested mechanism is the chameleon; it must satisfy the following equation~\cite{Khoury2004}:
\begin{equation}
\nabla^{2} \phi = V_{,\phi} + \frac{\beta}{M_{\rm{Pl}}} \rho\,
\label{eq:cham_field}
\end{equation}
where $V$ is the potential of the scalar field, $\beta$ is the coupling to the matter, $\rho$ is the matter 
density and $M_{\rm{Pl}} =\sqrt{8 \pi G}$ is the Planck mass.
The chameleon field leads to the fifth force in the form of: 
\begin{equation}
F(\phi) = - \frac{\beta}{M_{\rm{Pl}}} \nabla \phi\,
\label{eq:5thforce}
\end{equation}
$f(R)$ models show a chameleon mechanism having a fixed value of the coupling $\displaystyle{\beta=\sqrt{\frac{1}{6}}}$. 
These models have an additional degree of freedom that mediates the fifth force \cite{sotiriou}:
\begin{equation}
f_{,\rm{R}}(z) = - \sqrt{\frac{2}{3}}\frac{\phi_{\infty}}{M_{\text{Pl}}}\,
\label{eq:f_R}
\end{equation}
where $\phi_{\infty}$ describes the efficiency of the screening mechanism when $r\rightarrow \infty$. 
Thus, the $f(R)$ Lagrangian cannot be considered as a completely free function; it must satisfy many different 
constraints coming from both theoretical and phenomenological side. For example, in a region with high curvature, 
the first derivative of the Lagrangian must be $f_{,R}>0$, as well as the second derivative (we define: 
$f_{,RR}=d^2f(R)/dR^2$) $f_{,RR}>0$. These conditions will avoid tachyons and ghost solutions. Furthermore, to satisfy, 
Solar System constraints must be $|f_{,R0}|<<1$ in the present Universe. 
Two well-studied choices matching these constraints are the Starobinsky \cite{Starobroc} and Hu--Sawicki \cite{husawikbroc} 
models. In this review, we will focus only on the latter, which is expressed by the following Lagrangian:
\begin{eqnarray}
f(R)=-m^2\frac{c_1\left(\frac{R}{m^2}\right)^2}{c_2\left(\frac{R}{m^2}\right)^n+1}\,
\end{eqnarray}
from which one can obtain:
 \begin{eqnarray}
f_{,R} &=& -\frac{n{c_{1}}}{c_{2}^{2}} \left( {\frac{m^{2}}{R}} \right)^{n+1} \,
\label{eqn:minimum}
\end{eqnarray}
where the Ricci scalar for the flat $\Lambda$CDM model is:
\begin{equation}
R \approx 3 m^2 \left( a^{-3} + 4{\frac{\tilde\Omega_\Lambda}{\tilde\Omega_m}} \right) \,
\end{equation}

Here, the term $\displaystyle{\frac{\tilde\Omega_\Lambda}{\tilde\Omega_m}}$ matches the standard 
$\displaystyle{\frac{\Omega_\Lambda}{\Omega_m}}$ when $c_1 /c_2^2\rightarrow 0$. Hu and Sawicki \cite{husawikbroc} have 
shown that to recover GR results within the Solar System, one must be $|f,_{R0}|<10^{-6}$. 

\subsection{Analytical f(R) Gravity Models and Yukawa-Like Gravitational Potentials.}
\label{sub:yuka}

In this subsection, we will illustrate how an analytic $f(R)$ model expandable in Taylor series gives rise to a 
Yukawa correction to the Newtonian gravitational potential. Phenomenologically, this kind of correction 
behaves as the so-called chameleon mechanism, becoming negligible at small \linebreak scales \cite{annalen, chameleon}. In the case of the
post-Newtonian corrections, such a solution leads to 
redefining the coupling constants in order to fulfill the
experimental observations. In order to derive the correction terms to the Newtonian potential coming 
from an analytic $f(R)$ model, one must consider both the second and the fourth order in
the perturbation expansion of the metric.
Let us consider the
perturbed metric with respect to a Minkowskian background
$g_{\mu\nu}\,=\,\eta_{\mu\nu}+h_{\mu\nu}$. The metric components can
be developed as~follows:
\begin{eqnarray}\label{definexpans}
\left\{\begin{array}{ll} g_{tt}(t,
r)\simeq-1+g^{(2)}_{tt}(t,r)+g^{(4)}_{tt}(t,r)\,
\\\\
g_{rr}(t,r)\simeq1+g^{(2)}_{rr}(t,r)\,\\\\
g_{\theta\theta}(t,r)=r^2\,\\\\
g_{\phi\phi}(t,r)=r^2\sin^2\theta\,
\end{array}\right.\,
\end{eqnarray}
where $c\,=\,1$\,,\ $x^0=ct\rightarrow t$, and a general spherically-symmetric metric has been considered:
\begin{eqnarray}\label{me}
ds^2\,=\,g_{\sigma\tau}dx^\sigma
dx^\tau=-g_{00}(x^0,r)d{x^0}^2+g_{rr}(x^0,r)dr^2+r^2d\Omega\,
\end{eqnarray}
where $d\Omega$ is the solid angle. The next step is to introduce
an analytic Taylor expandable $f(R)$ functions with respect
to a certain value $R\,=\,R_0$:
\begin{eqnarray}\label{sertay}
f(R)=\sum_{n}\frac{f^n(R_0)}{n!}(R-R_0)^n\simeq
f_0+f_{,R0}R+f_{,RR0}R^2+f_{,RRR0}R^3+...\,
\end{eqnarray}

In order to obtain the weak limit, we must introduce Equations~(\ref{definexpans}) and (\ref{sertay}) in the 
field Equations~(\ref{field_eq}) and (\ref{trace_field_eq}) and to expand the equations up to the orders 
${\mathcal O}(0)$, ${\mathcal O}(2)$ and ${\mathcal O}(4)$.
 This approach permits us to select the Taylor coefficients in Equation~(\ref{sertay}). We remember that at zero order
${\mathcal O}(0)$, the~field equations give the condition
$f_0 =0$, and thus, the solutions at further orders do not depend on
this parameter. The field equations at ${\mathcal O}(2)$-order become:
\begin{eqnarray}\label{eq2}
\left\{\begin{array}{ll}
f_{,R0}rR^{(2)}-2f_{,R0}g^{(2)}_{tt,r}+8f_{,RR0}R^{(2)}_{,r}-f_{,R0}rg^{(2)}_{tt,rr}+4f_{,RR0}rR^{(2)}=0\,
\\\\
f_{,R0}rR^{(2)}-2f_{,R0}g^{(2)}_{rr,r}+8f_{,RR0}R^{(2)}_{,r}-f_{,R0}rg^{(2)}_{tt,rr}=0\,
\\\\
2f_{,R0}g^{(2)}_{rr}-r\left[f_{,R0}rR^{(2)}-f_{,R0}g^{(2)}_{tt,r}-f_{,R0}g^{(2)}_{rr,r}+4f_{,RR0}R^{(2)}_{,r}+4f_{,RR0}rR^{(2)}_{,rr}\right]=0\,
\\\\
f_{,R0}rR^{(2)}+6f_{,RR0}\left[2R^{(2)}_{,r}+rR^{(2)}_{,rr}\right]=0\,
\\\\
2g^{(2)}_{rr}+r\left[2g^{(2)}_{tt,r}-rR^{(2)}+2g^{(2)}_{rr,r}+rg^{(2)}_{tt,rr}\right]=0\,
\end{array} \right.\end{eqnarray}

As we can see, the fourth equation in the above system (the trace) gives us a differential equation with respect to the
Ricci scalar, which allows solving the system exactly
at ${\mathcal O}(2)$-order. 
Then, we obtain: 
\begin{eqnarray}\label{sol}
\left\{\begin{array}{ll}
g^{(2)}_{tt}=\delta_0-\dfrac{\delta_1}{f_{,R0}r}+\dfrac{\delta_2(t)}{3L}\dfrac{e^{-L
r}}{L r}+\dfrac{\delta_3(t)}{6L^2}\frac{e^{L
r}}{L r}\,
\\\\
g^{(2)}_{rr}=-\dfrac{\delta_1}{f_{,R0}r}-\dfrac{\delta_2(t)}{3L}\dfrac{L
r+1}{L r}e^{-L
r}+\dfrac{\delta_3(t)}{6\lambda^2}\dfrac{L r-1}{L
r}e^{L r}\,
\\\\
R^{(2)}=\delta_2(t)\dfrac{e^{-L
r}}{r}+\dfrac{\delta_3(t)}{2L}\dfrac{e^{L r}}{r}\,
\end{array}
\right.
\end{eqnarray}
where $L\,\doteq\,\sqrt{-\frac{f_{,R0}}{6f_{,RR0}}}$, $f_{,R0}$ and $f_{,RR0}$. In the limit $f(R)\rightarrow R$, we recover the standard Schwarzschild solution. Let us stress that $L$ has the dimension of $length$, the
integration constant $\delta_0$ is dimensionless, while $\delta_1(t)$ has the dimensions of $length^{-1}$ and $\delta_2(t)$ the dimensions of $length^{-2}$. Since the
differential equations in the system Equation~(\ref{eq2}) contain only spatial
derivatives, we can fix the functions of time
$\delta_i(t)$ ($i=1,2$) to arbitrarily-constant values. Furthermore, we can set $\delta_0$ to zero, because it is not an essential additive quantity. If we want a general solution of previous Equation~(\ref{sol}), we exclude the Yukawa growing mode, obtaining the following relations:
\begin{eqnarray}\label{mesol}
\left\{\begin{array}{ll}ds^2\,=\,-\biggl[1-\dfrac{r_g}{f_{,R0}r}+\dfrac{\delta_2(t)}{3\,L}\dfrac{e^{-L\,
r}}{L\,
r}\biggr]dt^2+\biggl[1+\dfrac{r_g}{f_{,R0}r}+\dfrac{\delta_2(t)}{3\,L}\dfrac{L\,
r+1}{L\, r}e^{-L\,
r}\biggr]dr^2+r^2d\Omega\,\\\\R\,=\,\dfrac{\delta_2(t)e^{-L\,
r}}{r}\,\end{array}\right.
\end{eqnarray}
where $r_g=2MG$.
Now, we can look for the solution in terms of the
gravitational potential from Equation~(\ref{sol}), obtaining the following relation:
\begin{eqnarray}\label{gravpot}
\Phi\,=\,-\frac{GM}{f_{,R0}r}+\frac{\delta_2(t)}{6\,L}\frac{e^{-L\,
r}}{L\, r}\,
\end{eqnarray}

Let us note that the $L$ parameter is related to the effective mass:
\begin{eqnarray}
m = \left(
\frac{-3}{L^2}\right)^{-\frac{1}{2}}\,=\,\left(\frac{2f_{,RR0}}{f_{,R0}}\right)^{\frac{1}{2}} 
\end{eqnarray}
and it can be interpreted also as an effective length.
Being $\displaystyle{1+\delta=f_{,R0}}$ and $\displaystyle{\delta_1=-\frac{6GM}{L^2}\frac{\delta}{1+\delta}}$ 
quasi-constant, the Equation~(\ref{gravpot}) becomes: 
\begin{equation}
\label{gravpot1} \Phi(r) = -\frac{G M}{
(1+\delta) r}\left(1+\delta e^{-\frac{r}{L}}\right)\,
\end{equation}

Here, the first term is the Newtonian-like part of the gravitational potential
to point-like mass, and the second
term is the Yukawa correction including a scale
length, $L$. If $\delta=0$, the Newtonian potential is recovered. 
 With this assumption, the new gravitational scale length $L$ could naturally arise and accommodate 
 several phenomena ranging from Solar System to cosmological scales. 

\section{Constraining \emph{f(R)} Gravity Models Using Clusters of Galaxies}
\label{sec.3}

Clusters of galaxies are the largest virialized object in the Universe. They are
the intermediate step between the galactic and the cosmological scales; thus, 
any relativistic theory of gravity must be capable of correctly describing their physics. 
Clusters typically contain a number of galaxies ranging from a few hundreds to one thousand,
grouped in a region of $\sim$2 Mpc, contributing 3\% to the total mass of the cluster. 
A more important component is represented by the baryons
residing in a hot inter cluster (IC) gas. Although IC gas is highly rarefied,
the electron number density is $n_e \sim 10^{-4}-10^{-2} \rm{cm}^{-3}$;
it makes up 12\% of the total mass; and it reaches high temperatures ranging from $10^7$ to $10^8$ K,
becoming a strong X-ray source with a luminosity typically ranging $L_X \sim 10^{43}-10^{45} {\rm erg/s}$.

One of the most promising tools to study clusters of galaxies is 
the Sunyaev--Zeldovich (SZ) \mbox{effect~\cite{SZ1972, SZ1980}}. CMB photons cross clusters of galaxies,
and they are scattered off by free electrons present in the hot IC gas. 
This interaction produces secondary temperature fluctuations of the CMB power spectrum due to: (1) 
the thermal motion of the electrons in the gravitational potential well of the cluster (it is named  
Thermal Sunyaev-Zeldovich (TSZ) 
; \cite{SZ1972}); (2) the kinematic motion of
the cluster as a whole with respect to the CMB rest
frame (it is named  
Kinetic Sunyaev-Zeldovich KSZ 
; \cite{SZ1980}). In the direction
of a cluster ($\hat{n}$), the SZ effect is given by:
\begin{equation}\label{eq:sz}
\frac{T(\hat{n})-T_0}{T_0}=\int \left[g(\nu)\frac{k_BT_e}{m_ec^2}+
\frac{\vec{v}_{cl}\hat{n}}{c}\right]d\tau
\end{equation}
where $d\tau=\sigma_Tn_edl$ is the optical depth, 
$\sigma_T$ is the Thomson cross-section, 
$k_B$ is the Boltzmann constant, $m_ec^2$ is the electron annihilation temperature, $c$ is
the speed of light, $\nu$ is the frequency of the observation and $\vec{v}_{cl}$
is the peculiar velocity of the cluster. $T_0=2.725\pm 0.002$ K \cite{Fixsen2009} is the current 
CMB blackbody temperature, and $g(\nu)$ is the frequency dependence of the TSZ effect.
In the non-relativistic limit, $g(x)= x{\rm coth}(x/2)-4$, with
$x=h\nu/k_BT$ the reduced frequency. 
The physical description
of the TSZ is commonly given by introducing the Comptonization parameter:
\begin{equation}\label{eq:y_c}
y_c=\frac{k_B\sigma_T}{m_ec^2}\int n_e(r)T_e(r) dl=
\frac{\sigma_T}{m_ec^2}\int P_e(r) dl
\end{equation}
where $P_e(r)$ is the electron pressure profile that must be specified to 
predict the TSZ anisotropies. 
Using X-ray observations and numerical simulations, several 
cluster profiles $\left(n_eT_e\right)$ have been proposed.

Isothermal $\beta$-model fits the X-ray emitting region of the clusters of galaxies well 
 \cite{cavaliere1976, cavaliere1978}, with the electron density given by:
 \begin{equation}
 n_e(r)=n_{e,0}\left[1+\left(\frac{r}{r_c}\right)^2\right]^{-\frac{3\beta}{2}}
 \end{equation}
 where $n_{e,0}$ is the central electron density and $r_c$ is the core radius.
 Those parameters, together with the electron temperature and the slope $\beta$,
 are determined from observations. Using the X-ray surface brightness of clusters, the slope 
 value ranges in the interval $[0.6-0.8]$ \cite{jones1984}.
 
 The isothermal $\beta$-model over-predicts
the TSZ effect in the outskirts of the cluster of galaxies \cite{atrio2008}.
Recently, a phenomenological parametrization of the electron pressure profile, derived from the numerical simulations,
has been proposed \cite{nagai2007, arnaud2010}. The functional form of  Universal pressure profile is: 
\begin{equation}
p(x) \equiv \frac{P_0}{(c_{500}x)^{\gamma_a} 
[1+(c_{500}x)^{\alpha_a}]^{(\beta_a-\gamma_a)/\alpha_a}}
\label{eq:universal_profile}
\end{equation}
where $x=r/r_{500}$, and $r_{500}$ is the radius at which the mean overdensity of 
the cluster is $500$-times the critical density of the Universe at the
same redshift. 
Then, $c_{500}$ is the concentration parameter at $r_{500}$.
The model parameters were constrained using X-ray data \cite{arnaud2010, Sayers2012} or
CMB data \cite{Planck_int_5}; their best fit values are quoted in Table~\ref{table1}.


\begin{table}[H]
\centering
\small
\begin{tabular}{ccccccc}
\toprule
\textbf{Model} & \boldmath$c_{500}$ &\boldmath $\alpha_a$ & \boldmath$\beta_a$ & \boldmath$\gamma_a$ & \boldmath$P_0$ & {\bf Reference}\\ 
\midrule
Arnaud \emph{et al.} 2010 & 1.177 & 1.051 & 5.4905 & 0.3081 & $8.403$ $h_{70}^{3/2}$ & \cite{arnaud2010}\\ 
Sayers \emph{et al.} 2013 & 1.18 & 0.86 & 3.67 & 0.67 & 4.29 & \cite{Sayers2012}\\
Planck \emph{et al.} 2013 & 1.81 & 1.33 & 4.13 & 0.31 & 6.41 & \cite{Planck_int_5}\\
\bottomrule
\end{tabular}
\caption{Parameters of universal pressure profile fitted by different groups
using \mbox{X-ray~\cite{arnaud2010, Sayers2012}} and CMB data \cite{Planck_int_5}.}
\label{table1}

\end{table}

Since TSZ does not depend on redshift in the $\Lambda$CDM model,
clusters are a very important laboratory to test cosmology (see \cite{Allen2011, Mana2013, Planck2015_24, Sartoris2015} 
and the references within),
even to test the fundamental pillars of the Big Bang scenario \cite{DeMartino2012, DeMartino2015b}.

In the last decade, SZ multi-frequency measurements have been
reported by several groups: the~Atacama Cosmology Telescope (ACT) 
\cite{Hand2011,Sehgal2011,Hasselfield2013,Menanteau2013}, 
the South Pole Telescope (SPT) 
\cite{Staniszewski2009,Vanderlinde2010,Williamson2011,Benson2013}
and the Planck satellite \cite{Planck_int_5, Planck_int_10, Planck2015_22, Planck2015_27}.
 Wilkinson Microwave Anisotropy Probe (WMAP)
 seven years \cite{Komatsu2011} 
and Planck \cite{Planck_int_5, Planck_int_10} 
data have discussed several apparent discrepancies with 
$\Lambda$CDM predictions. If these discrepancies
are due to the physical complexity of clusters of galaxies or are due to
the limitation of the theoretical modeling \cite{Fusco2012}, it~is an open~question.

To study if these discrepancies are due to the theoretical modeling and, at the same time,
to constrain/rule out ETGs at cluster scales, the pressure profile of a cluster of galaxies has been considered.
On the one hand, it is interesting to study models that can explain the cluster without requiring 
any dark component \cite{DeMartino2014}. On the other hand, models constructed to 
mimic the $\Lambda$CDM expansion history by replacing the DE with higher order terms in the gravitational Lagrangian must also 
describe the emergence of the LSS and the dynamical properties of clusters 
\cite{Terukina2012, Terukina2014, Wilcox2015}.

\subsection{Pressure Profile from Yukawa-Like Gravitational Potential.}

One approach to test ETGs using the TSZ pressure profile is focused on the $f(R)$ Lagrangian that is expandable 
in Taylor's series (Equation~\eqref{sertay} in Section \ref{sub:yuka}), where the Yukawa-like correction to the Newtonian potential, Equation~\eqref{gravpot1}, could be 
interpreted as the dynamical effect of DM in clusters of galaxies~\cite{DeMartino2014}.
Making the hypothesis that: the gas is in hydrostatic equilibrium within 
the modified gravitational potential well (without any DM term):
\begin{equation}\label{eq:HE}
  \frac{dP(r)}{dr}=-\rho(r)\frac{d\Phi(r)}{dr}\,
\end{equation}
the gas follows a polytropic equation of state: 
\begin{equation}\label{eq:PES}
 P(r)\propto\rho^\gamma(r)\,
\end{equation}
the Equations~\eqref{gravpot1}, \eqref{eq:HE} and \eqref{eq:PES} can be numerically integrated
to compute the pressure profiles by closing the system with the equation for the mass conservation:
\begin{equation}\label{eq:EMC}
  \frac{dM(r)}{dr}= 4\pi\rho(r)
\end{equation}

Thus, the pressure profile will be a function of 
the two extra gravitational parameters $(\delta, L)$ 
and the polytropic index $\gamma$. Let us remark that the density $\rho(r)$ includes only
baryonic matter, without resorting to DM particles.

\subsubsection{Data and Results.}

\begin{figure}[H]
\centering 
\includegraphics[width=0.69\columnwidth]{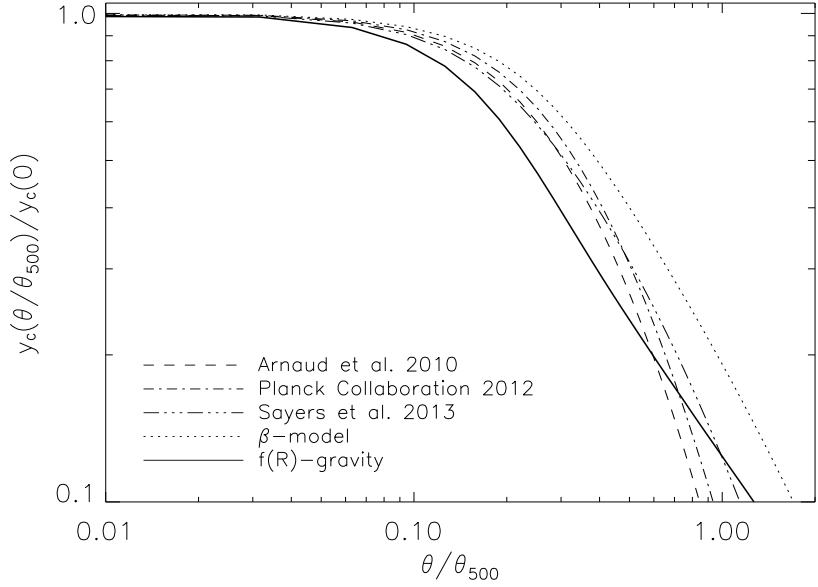}\\
\caption{Comptonization parameter for Coma
 cluster ($z=0.023$).
The pressure profile is integrated along the line of sight for:
the three universal profiles (dashed, solid and dash-dotted lines; 
the model parameters are quoted in Table~\ref{table1}); 
$\beta=2/3$ model (long dashed line); and the $f(R)$ model (red solid line, 
$[\delta, L, \gamma]=[-0.98, 0.1, 1.2]$).\vspace{-12pt}}\label{fig:Fig1}
\end{figure}

To constrain the model, the {Planck} (all Planck data are publicly available and can be downloaded 
from {http://www.cosmos.esa.int/web/planck}) 2013 data and 
an X-ray clusters catalog \cite{Kocevski2006} have been used. 
In~Figure~\ref{fig:Fig1}, the predicted Comptonization parameter for the universal pressure
profile with the parameters given in Table \ref{table1} and the $\beta=2/3$-model is compared
with the $f(R)$-model with the parameters $[\delta, L, \gamma]=[-0.98, 0.1, 1.2]$.
The model is particularized for the Coma
 cluster, located at redshift $z=0.023$,
with core radius $r_c=0.25$ Mpc, X-ray temperature $T_X=6.48$ keV and 
electron density $n_{e,0}=3860\,\rm{m}^{-3}$. The plot shows 
that the higher order terms in the gravitational $f(R)$-Lagrangian 
could potentially explain the cluster without accounting for DM 
contribution. 
 
The SZ emission was measured on the Spectral Matching Independent Component
Analysis (SMICA) 
 map (SMICA is a foreground-cleaned map; it 
has been constructed using a component separation method by combining the data at all frequencies
\cite{Planck2013_12}; the~SMICA map has a $5'$ resolution) at the locations of all X-ray selected clusters,
and then, it was compared with the model prediction. The temperature anisotropies
were averaged over rings of width $\theta_{500}/2$, where $\theta_{500}$ is 
the angular scale subtended by the $r_{500}$. The error associated with each data point was computed carrying 
out $1000$ random simulations and evaluating the profile for
each simulation. The~analysis was performed by computing
the likelihood function $\log{\cal L}=-\chi^2/2$ as:
\begin{equation}
\chi^2 ({ {\bf p}})=\Sigma_{i,j=0}^{N} 
(y({ {\bf p}}, x_i)-d(x_i))C_{ij}^{-1}
(y({ {\bf p}}, x_j)-d(x_j))\,
\label{eq:chi}
\end{equation}
where $N$ is the number of data points, and
${{\bf p}}=(\delta,L,\gamma)$. In Equation~(\ref{eq:chi}), $d(x_i)$ are the data,
and $C_{i,j}$ is the correlation matrix between the average
temperature anisotropy on the discs and rings. 
Two~parameterizations have been tested: 
(A) $L= \zeta r_{500}$, the scale length is different for each; clusters scale linearly with $r_{500}$; 
(B) $L$ has the same value for all of the clusters.
The models were constructing choosing ``physical'' flat priors on the parameters. Looking at the Yukawa-potential in 
Equation~\eqref{gravpot1}, if $\delta<-1$, it becomes repulsive, and if 
$\delta=-1$, it diverges; while the polytropic index $\gamma$ was varied
in the range corresponding to an isothermal and an adiabatic state of a 
monoatomic gas, respectively. The~{priors} are summarized in Table \ref{tab:table2}.

\begin{table}[H]
\centering
\small
\begin{tabular}{ccccc}
\toprule
\textbf{Parameterization} & \boldmath$\delta$ & \boldmath$\gamma$ & \boldmath$L$ & \boldmath$\zeta$ \\ 
\midrule
(A) & $[-0.99,1.0]$ & $[1.0,1.6]$ & - & $[0.1,4]$\\
(B) & $[-0.99,1.0]$ & $[1.0,1.6]$ & $[0.1,20]$& -\\
\bottomrule
\end{tabular}
\caption{{Priors} on the parameters when modeling the cluster profile in $f(R)$-gravity.}
\label{tab:table2}

\end{table}

\begin{figure}[H]
\centering 
\includegraphics[width=0.95\columnwidth]{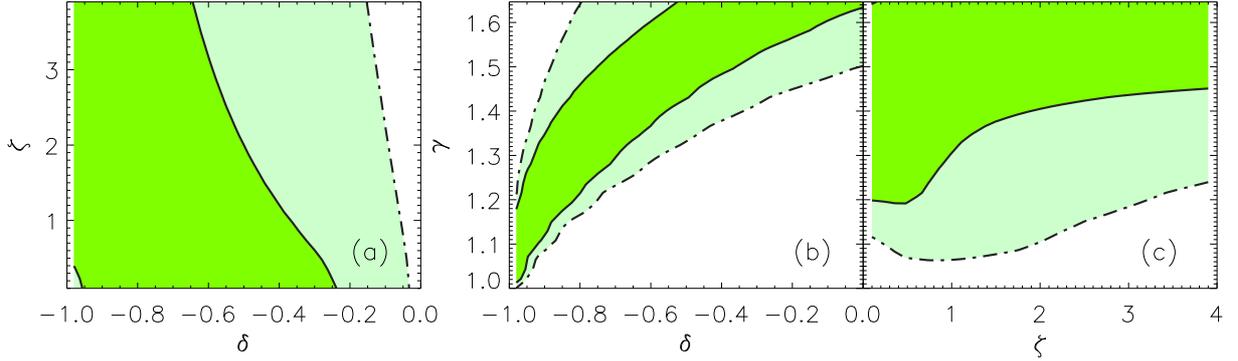}\\
\caption{2D contours at the 68\% (dark green) and 95\% (light green) confidence levels of the marginalized likelihoods 
for Parameterization (A). In panel (a), (b), and (c) there are shown the 2D contours for the parameters ($\zeta, L$), 
($\delta, L$), and ($\delta, \zeta$), respectively. Since the contours are opened, only upper limits on the parameters can 
be given. 
\vspace{-12pt}}\label{fig:Fig2}
\end{figure}

\begin{figure}[H]
\centering 
\includegraphics[width=0.95\columnwidth]{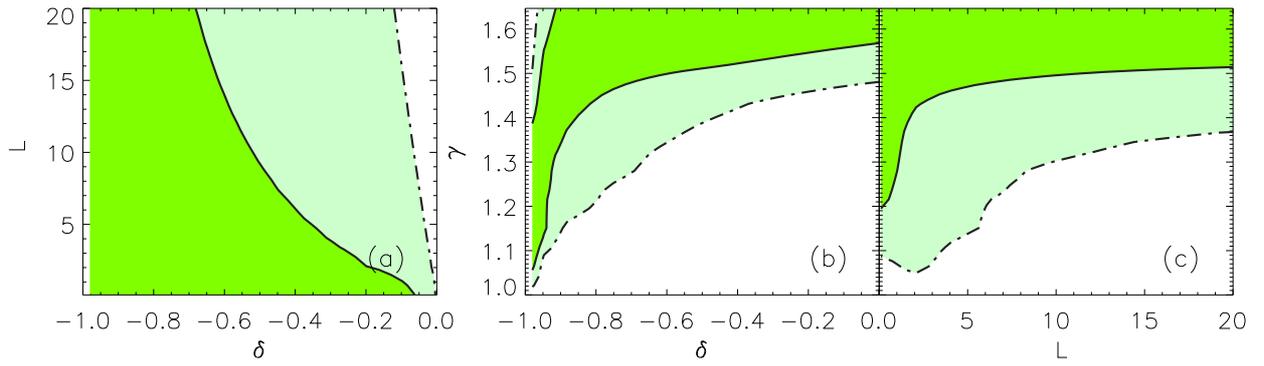}\\
\caption{2D contours from the marginalized likelihoods 
for Parameterization (B). Contours follow the same convention of Figure \ref{fig:Fig2}. 
This parameterization also provides opened contours. Therefore, also in this case, one can only give upper
limits on the parameters and can not distinguish which parameterization is the best one.\vspace{-6pt}}\label{fig:Fig3}
\end{figure}

Figures~\ref{fig:Fig2} and \ref{fig:Fig3} show the 2D contours at 
the $68\%$ and $95\%$ confidence levels of the marginalized likelihoods 
for both Parameterizations (A) and (B), respectively. 
Although the contours are opened and only upper limits can be 
obtained, the value $\delta=0$ is {always excluded} at more than the $95\%$ confidence level.
Therefore, Newtonian potential without DM cannot fit cluster pressure profile, and 
thus, {either DM or modified gravity must be the right description of the dynamical properties
of the cluster of galaxies}. The~upper limits are summarized in Table \ref{tab:table3}.

\begin{table}[H]
\centering
\small
\begin{tabular}{ccccc}
\toprule

   & \textbf{68\% CL} & \textbf{95\% CL} & \textbf{68\% CL} & \textbf{95\% CL} \\ 
\midrule
$\delta$ & $<$$-0.46$ & $<$$-0.10$ & $<$$-0.43$ & $<$$-0.08$ \\
$\gamma$ & $>$$1.35$& $>$$1.12$ & $>$$1.45$& $>$$1.2$ \\
$L (or \, \zeta)$ & $<$$2.5$ & $<$$3.7$ & $<$$12$ & $<$$19$ \\
\bottomrule
\end{tabular}
\caption{Upper limits on the model parameters for the cluster profile in $f(R)$-gravity.}
\label{tab:table3}

\end{table}

\subsection{Chameleon Gravity: Hydrostatic and Weak Lensing Mass Profile of Galaxy Cluster.}
\label{}

A second approach in using clusters of galaxies to test ETGs is focused on the
chameleon gravity models that introduce a scalar field
non-minimally coupled to the matter components. This coupling gives rise to the fifth
force, which is tightly constrained at Solar System scales (see Section \ref{sub:chama}). The~main difference with the other approach discussed above is that the chameleon gravity only eliminates DE, preserving the DM component
as the source of the gravitational field needed to explain the emergence of the LSS.

Because of the screening mechanism, the effect of this force is negligible at small scales, 
and thus, it does not affect the density distribution of the IC gas in the cores of galaxy clusters, 
but its effect may not be totally screened in the outskirts. As a consequence,
the gas distribution will be more compact in the outskirts under the effect of
an additional pressure term that balances the effect of the fifth force.
Thus, the~hydrostatic mass of a cluster should be different from the mass obtained from 
weak gravitational lensing that depends only on the distribution of DM along the line of sight
and does not assume hydrostatic~equilibrium.

The model studied in \cite{Terukina2012, Terukina2014} assumes the spherically-symmetric distribution of IC gas and DM.
The IC gas is in hydrostatic equilibrium within the potential well
generated by the DM distribution. Departures from hydrostatic equilibrium are parameterized,
introducing a non-thermal term in the pressure, while the chameleon field directly contributes 
to the total amount of mass. The model is compared to lensing observations used to estimate
the mass, with X-ray and SZ observations used to reconstruct the gas pressure
profile \cite{Terukina2012, Terukina2014, Wilcox2015}.

Let us start writing down the equation of the hydrostatic equilibrium for the IC gas component:
 \begin{eqnarray}\label{eq:he}
 \frac{1}{\rho_{\rm gas}(r)}\frac{dP_{\rm tot}(r)}{dr} =-\frac{GM(<r)}{r^2}
 \end{eqnarray}
 where $ M_{\rm}(<r)$ is the total mass enclosed in a sphere of radius $r$, 
 $\rho_{\rm gas}$ is the gas density distribution within the sphere and $P_{\rm tot}$ 
 is the total gas pressure that can be recast as:
 \begin{equation}
 P_{\rm tot}(r)=P_{\rm th}(r)+P_{\rm non-th}(r)+P_\phi(r)
 \end{equation}
 including both thermal $(P_{\rm th}(r))$ and non-thermal $(P_{\rm non-th}(r))$ pressure, and the term due to the chameleon field $\left(P_\phi(r)\right)$. Thus,
 the total mass in Equation~\eqref{eq:he} is:
 \begin{eqnarray}\label{eq:HE1}
 M_{\rm}(<r)&=&M_{\rm th}(r)+M_{\rm non-th}(r)+M_\phi(r)
 \end{eqnarray}
 
 By definition, each term can be expressed as:
 \begin{eqnarray}
 && M_{\rm th}(r)\equiv-\frac{r^2}{G\rho_{\rm gas}(r)}\frac{dP_{\rm th}(r)}{dr}
\label{eq:mrrp}
\\
 && M_{\rm non-th}(r)\equiv-\frac{r^2}{G\rho_{\rm gas}(r)}\frac{dP_{\rm non-th}(r)}{dr} \label{eq:nothermalmass}\\
 && M_{\phi}(r)\equiv-\frac{r^2}{G}{\frac{\beta}{M_{\rm Pl}}}{d\phi(r)\over dr}
 \label{eq:chameleonmass}
\end{eqnarray}
$M_\phi(r)$ is the mass associated with the chameleon field, and $M_{\rm non-th}$ represents the departure 
from the hydrostatic equilibrium mass $M_{\rm th}$. Introducing the equation of state of the IC gas, 
\begin{equation}
P_{\rm th}(r)=kn_{\rm gas}(r)T_{\rm gas}(r)
\end{equation}
the Equation~\eqref{eq:mrrp} can be re-written as
\begin{eqnarray}
&& M_{\rm th}(r)=-\frac{kT_{\rm gas}(r)r}{\mu m_{\rm p}G}
\left(\frac{d\ln \rho_{\rm gas}(r)}{d\ln r}+\frac{d\ln T_{\rm gas}(r)}{d\ln r}\right)
\label{eq:mrrt}
\end{eqnarray}
where the identity $\rho_{\rm gas}(r)=\mu m_{\rm p}n_{\rm gas}(r)$ has been considered with the mean molecular weight $\mu$ and the proton mass $m_{\rm p}$.

According to hydrodynamical simulations of the $\Lambda$CDM model~\cite{Battaglia,Shaw}, 
the non-thermal pressure term can be modeled as: 
 \begin{eqnarray}
 P_{\rm non-th}(r)=\frac{g(r)}{1-g(r)}n_{\rm gas}(r)kT_{\rm gas}(r)
 \end{eqnarray}
with:
 \begin{eqnarray}
 \label{eq:grg}
 &&g(r)=\alpha_{\rm nt}(1+z)^{\beta_{\rm nt}}
 \biggl({r\over r_{500}}\biggr)^{n_{\rm nt}}\biggl({M_{200}\over
 3\times10^{14}M_\odot}\biggr)^{n_{\rm M}}
 \end{eqnarray}
where $\alpha_{\rm nt}$, $\beta_{\rm nt}$, $n_{\rm nt}$ and $n_{\rm M}$ are constants. Therefore,
the non-thermal mass term is:
\begin{equation}
 M_{\rm non-thermal}= -{r^2\over G \rho_{\rm gas}^{\rm (X)}}{d\over dr}\left(
{g\over 1-g}n_{\rm gas}^{\rm (X)}kT_{gas}^{\rm (X)}\right)
\label{eq:nonthmX}
\end{equation}

This term is modeled without considering the chameleon field, and it 
has been shown to reproduce accurately the non-thermal contribution, even
in $f(R)$-gravity. Nevertheless, a more accurate 
procedure would require modeling this term from hydrodynamical simulations
under chameleon gravity. Finally, the total mass inferred from hydrostatic equilibrium has to be equal to the 
the total mass as inferred from weak lensing (WL):
\begin{equation}
 M_{\rm tot} = M_{\rm thermal}+M_{\rm non-thermal}+M_\phi\equiv M_{\rm WL}
\end{equation}
 
The DM density distribution is equally well described in the $f(R)$-model and in Newtonian gravity
using the Navarro--Frenk--White (NFW) profile \cite{Lombriser2012a}. This profile has 
the following functional form~\cite{NFWProfile}:
 \begin{eqnarray}
 \rho(r)=\frac{\rho_{\rm s}}{r/r_{\rm s}\left(1+r/r_{\rm s}\right)^2}\, \label{eq:nfwfit}
 \end{eqnarray}
and it has been constructed using $N$-body simulations of the $\Lambda$CDM model. 
Thus, the DM mass within the radius $r$ is: 
 \begin{eqnarray}\label{eq:NFWmass0}
 M_{\rm DM}(<r)=4\pi \int_0^r dr r^2\rho(r) =4\pi \rho_{\rm s}r_{\rm s}^3 \left[\ln(1+r/r_{\rm s})-\frac{r/r_{\rm s}}{1+r/r_{\rm s}}\right]\equiv M_{\rm WL}
\end{eqnarray}
that is equivalent to the WL mass if the DM component dominates over the baryonic one.
The model has been tested using only COMA cluster \cite{Terukina2014} and using a sample
of 58 X-ray selected cluster associated also with weak lensing measurements \cite{Wilcox2015}.

\subsubsection{{Data and Results}}

The chameleon model is essentially described by the following parameters $(\beta, \phi_\infty)$
in Equations~\eqref{eq:cham_field} and \eqref{eq:f_R}.
In order to normalize them, those parameter have been replaced by:
\begin{align}
 & \beta_2= \frac{\beta}{1+\beta},\\
 & \phi_{\infty,2}=1- e^{-\frac{\phi_{\infty}}{10^{-4}M_{\rm Pl}}} 
\end{align}
and constrained by carrying out the Monte Carlo Markov chain (MCMC) method using X-ray, SZ 
and WL observations. Another important parameter for studying clusters of galaxies is the critical radius:
\begin{equation}
r_{\rm{crit}} = \frac{\beta \rho_{\rm{s}} r^{3}_{\rm{s}}}{M_{\rm{Pl}} \phi_{\rm{\infty}}} - r_{\rm{s}}
\end{equation}
where $\rho_{\rm{s}}$ is the density at this radius. It represents the distance from the DM halo center 
where the screening mechanism acts and it is not possible to
distinguish between chameleon gravity and GR \cite{Terukina2012}.

\begin{figure}[H]
\centering
\includegraphics[width=11cm]{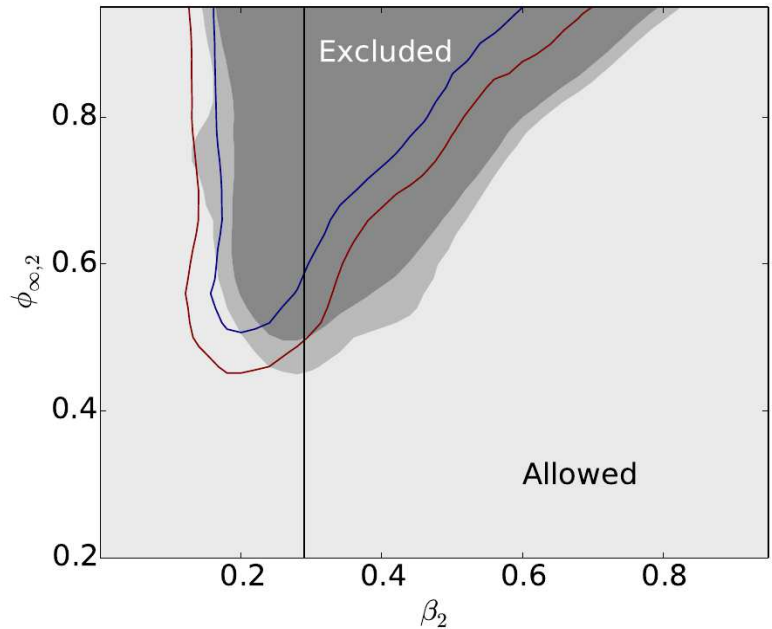}
\caption{The confidence contours for the renormalized parameters $(\phi_{\infty \rm{,2}},\beta_{\rm{2}})$
The $95\%$ and $99\%$ confidence levels are plotted in light gray and medium gray, respectively. The results are from
\cite{Wilcox2015}. The $95\%$ and $99\%$ confidence contours from \cite{Terukina2014} are over-plotted also in
red and blue, respectively. The vertical line corresponds to $|f_{\rm{,R0}}| < 6 \times 10^{-5}$.\vspace{-5pt}}
\label{fig:MG_contours}
\end{figure}

In \cite{Terukina2014}, the authors have constrained chameleon gravity only using the Coma cluster ($z=0.0231$). 
For this cluster, they implement the MCMC method using
the X-ray temperature profile reported by the X-ray Multi-Mirror Mission {(XMM)
-Newton} \cite{Coma_3} and {Suzaku} \cite{Coma_4}
for the inner and outer region, respectively. Then,
the X-ray surface brightness profile was taken from {XMM-Newton} \cite{Coma_2}
and the SZ pressure profile measured by {Planck} \cite{Planck_int_10}. For the WL counterpart,
the measurements were reported in \cite{Coma_7}. More recently, in~\cite{Wilcox2015},
the authors have constrained the chameleon model using publicly-available weak lensing data provided
by the Canada France Hawaii Lensing Survey (CFHTLenS; \cite{Heymans2012}),
and X-ray data taken from the {XMM-Newton} archive. Their sample includes 
58 clusters spanning the redshift range $z=[0.1-1.2]$, with measured X-ray temperatures 
$T_{\rm{x}}=[0.2 - 8]$~keV. For this sample of clusters, they also carried out an MCMC analysis.
In Figure \ref{fig:MG_contours} are shown the two-dimensional contours for the parameters
$\beta_{\rm{2}}$ and $\phi_{\infty \rm{,2}}$ as generated by \cite{Wilcox2015}.
The red and blue lines are the $95\%$ and $99\%$ confidence limits from~\cite{Terukina2014}.
Those results represent the best constraints on chameleon gravity using clusters. On~the one hand,
at low values of $\beta$, the departure from GR is too small to be
detected with these observational errors. On the~other hand, GR gravity is recovered 
within the critical radius $r_{\rm{crit}}$, and large values of $\beta$ lead 
to a lower value for $\phi_{\infty}$ that could keep $r_{\rm{crit}}$ in the whole 
cluster region. Although
the analysis from \cite{Wilcox2015} using the profiles outside of the critical radius of the cluster
could better constrain large values of $\beta$,
the analysis carried out by \cite{Terukina2014} also used SZ measurements and could discriminate better between
GR and chameleon gravity at lower values of $\beta$.
In both \cite{Terukina2014,Wilcox2015}, it also constrained the term $f_{\rm{,R0}}$ from 
Equation~\eqref{eq:f_R}. The~result provides an upper limit today $|f_{\rm{,R0}}| < 6 \times 10^{-5}$ 
at a $95\%$ of confidence level.
\newpage
\section{N-Body Hydrodynamical Simulations in {\it f(R)} Gravity}
\label{sec.4}

All chameleon-like theories of gravity need a specific screening mechanisms (Section \ref{sub:chama}).
The effect of all such mechanisms is reflected on the non-linear
perturbations of the matter distribution. Therefore, some of the most optimal tools
to provide a full description of these mechanisms are 
modifications of standard N-body algorithms 
\cite{Oyaizu2008, Khoury2009, Li2012b, Brax2011, Davis2012,Baldi2012, Llinares2013a}.

The aims were related to the study of the non-linear effects on the 
emergence of the LSS, such as 
the matter power spectrum \cite{Oyaizu2008, Li2012b, Schmidt2009}, the redshift-space
distortions \cite{Jennings2012} and the physical and statistical properties of the 
cold DM halos and voids \cite{Lombriser2012a, Lee2013, Lam2012, Llinares2013b, Zhao2010, Zhao2011, Winther2012, Li2012a}. 
However, the same scales and structures that could provide a powerful tool for
testing the theory of gravitation are also seriously affected by
non-linear astrophysical processes \cite{Puchwein2005, Puchwein2008,
Stanek2009, vanDaalen2011, Semboloni2011, Casarini2011}. 
In the few last years, a very powerful code
to investigate the combined effects of astrophysical processes and alternative theories
of gravity has been developed~\cite{Puchwein2013}. The~code is named Modified Gravity (MG)-GADGET.
It was used to study the Hu--Sawicki model \cite{husawikbroc},
pointing out the existence of an important degeneracy between
modified gravity models and the uncertainties in the baryonic physics
particularly, related to AGN feedback \cite{Puchwein2013}. 
Afterwards, the code was combined with a massive
neutrinos algorithm \cite{Viel2010}, to study the 
cosmic degeneracy between modified gravity and massive neutrinos.
It was found that the deviations from the matter power spectrum
in the $\Lambda$CDM model are, at most, of the order of 10\%
at $z=0$ taking $f(R)$ models with $f_{,R0}=-1\times10^{-4}$, and the total
neutrinos mass $\Sigma_i m_{\nu_i}=0.4$ eV \cite{Baldi2014}.
Using again the Hu--Sawicki parameterization, the effects of the 
model on the IC gas and its physical properties have been studied.
It was found that the DM velocity dispersions in the 
$f(R)$ model are $\sim$$4/3$ larger than in $\Lambda$CDM, when low mass halos 
are considered. Nevertheless, the mass of the objects that are affected by the chameleon
field depend on the value of the background field: for $|f_{,R0}|=10^{-4}$ the field
is never screened, nor for massive clusters; 
for $|f_{,R0}|=10^{-5}$, the chameleon field is screened for mass above $\sim$$10^{14.5} M\odot$;
and~for $|f_{,R0}|=10^{-6}$, the Hu--Sawicki model does not produce any effect in the range of 
mass explored in~\cite{Arnold2014}. Finally, MD-GADGET and the Hu--Sawicki model
have been also used to study the synthetic Lyman-$\alpha$ absorption spectra using simulations that
include star formation and cooling effects. The~discrepancies with the simulations based on the $\Lambda$CDM
model were at most $10\%$ with the background field fixed at $|f_{,R0}|=10^{-4}$ \cite{Arnold2015}.

All tests carried out using N-body simulations have provided several indications about the presence
of a degeneracy between the effects of the complex baryonic physics (highly non-linear) and the effects
of the modifications of the theory of gravitation. Even with the high accuracy of future data, without
a full understanding of the non-linear astrophysical processes, it would be not possible 
to break the degeneracy and to constrain or rule out ETGs. Therefore, there is the need to fully explore
this degeneracy with N-body simulations to know the level of bias introduced.

\section{Constraining the Expansion History of the Universe in {\em f(R)} Gravity}
\label{sec.5}

Cosmological models can be tested looking at background evolution and also
at the growth of the cosmic structures. Testing the $f(R)$ model at the background level
allows finding the so-called ``{realistic}'' models that have the radiation, matter and DE
eras very close to the $\Lambda$CDM ones. Although many {realistic} $f(R)$ models have
been constructed, it should be considered
that different models with a comparable late-time accelerated expansion
and similar background densities' evolution could differ {in the growing of the} matter density
perturbations \cite{Starobinsky1998}. {For example, it is well known that the effects at the background
level of considering a collisional matter component filling the Universe are negligible. Nevertheless,
non-collisional and collisional matter affect in different ways the evolution of matter perturbations, 
opening the possibility of distinguishing them \cite{Oikonomou2015}}.
Another problem is represented by the need 
to break the degeneracies between the evolving DE models and modified gravity in order to
confirm/rule out models \cite{Gong2008}. Thus, the growth factor (GF) data become extremely important
to test modified gravity and to find their signatures. However, the available data have less accuracy than the other 
datasets, such as SNeIa, Baryonic Acoustic Oscillation (BAO)
 and $H(z)$; thus, it is not possible to fully constrain $f(R)$-models. 
This scenario will totally change with the forthcoming Euclid observations. That dataset will have an 
unprecedented accuracy, and it will allow constraining deviations from the standard model at a few percent
level \cite{Euclid2013}.

Let us start by the flat Friedman--Lemaitre--Robertson--Walker metric (FLRW),
\begin{equation}\label{eq:FLRW}
 ds^2=-dt^2+a^2(t)dx^2
\end{equation}
where $a(t)$ is the rate expansion of the Universe. The Ricci scalar is related to the Hubble function,
$H\equiv\dfrac{\dot{a}(t)}{a(t)}$ by the following expression:
\begin{equation}\label{eq:Ricci}
 R=6\left(2H^2+\dot{H}\right)
\end{equation}
and using the non-relativistic matter and radiation approximation, the following conservation laws
are~satisfied:
\begin{align}\label{eq:conslaw}
 \dot{\rho}_m+3H\rho_m=0\\
 \dot{\rho}_\gamma+3H\rho_\gamma=0
\end{align}

Using Equations~\eqref{field_eq}, \eqref{trace_field_eq}, \eqref{eq:FLRW} and \eqref{eq:Ricci},
the modified Friedman equations are obtained \cite{hwang1991}:
\begin{align}
 3f_{,R}H^2=\kappa^2(\rho_m+\rho_\gamma)+\frac{1}{2}(f_{,R} R-f(R))-3H\dot{f}_{,R}\,\label{eq:mod_friedmann1}\\
 -2f_{,R}\dot{H}=\kappa^2(\rho_m+\frac{4}{3}\rho_\gamma)+\ddot{f}_{,R}-H\dot{f}_{,R}\, \label{eq:mod_friedmann2}
\end{align}

To compute the the evolution of the matter density perturbations, 
$\delta_\mathrm{m} \equiv \frac{\delta \rho_\mathrm{m}}{\rho_\mathrm{m}}$, one can mainly follow two different approach:

$\bullet$ {Sub-horizon approximation:} The evolution of the matter density perturbations 
can be obtained solving the following equation \cite{Boisseau_deltarho}: 
\begin{equation}\label{eq:iv1}
 \ddot{\delta}_\mathrm{m} \, + \, 2 H \dot{\delta}_\mathrm{m} \, - \, 4 \pi G_\mathrm{eff}(a,k) \rho_\mathrm{m} \delta_\mathrm{m} = 0
\end{equation}
where $k$ is the comoving wave number, and $G_\mathrm{eff}(a,k)$ is the effective gravitational constant.
The explicit forms of $G_{eff}$ depend on the ETGs model. In the case of $f(R)$ gravity, it is 
given by \cite{Tsujikawa:2007gd}:
\begin{equation}\label{eq:iv2}
 G_\mathrm{eff}(a,k) = \frac{G}{f_{,R}} \left[ 1 + \frac{\left( k^2/a^2 \right) \left( f_{,RR}/f_{,R} \right)}{1 + 3
 \left( k^2/a^2 \right) \left( f_{,RR}/f_{,R} \right)} \right]\,
\end{equation}

By definition, the growth factor $f_g$ is: 
\begin{equation}
f_g \equiv \frac{d\ln \delta_\mathrm{m}}{d \ln a}
\end{equation}
thus, Equation~\eqref{eq:iv1} becomes:
\begin{equation}\label{eq:deltagf}
\frac{df_g}{d\ln a}+f_g^2+\left(\frac{\dot{H}}{H}+2\right)f_g=\frac{3}{2} \frac{8\pi G_\mathrm{eff} \rho_\mathrm{m}}{3H^2}
\end{equation}

Since Equation~\eqref{eq:iv1} does not give raise to analytical solutions, it is customary to introduce an efficient
parameterization of the matter perturbations. The most common one is the so-called~{$\gamma$-parameterization}: 
\begin{equation}\label{eq:gamma_par}
f_g=\Omega_{m}^{\gamma}\,
\end{equation}
 $f_g^{\Lambda{\rm{CDM}}}$ is obtained for the value $\gamma\approx0.545$. Therefore,
any departures from this value could indicate the need to use a different model. More in general, 
the $\gamma$ function could evolve with the redshift in the case of 
modified gravity models \cite{gannouji09,fu09, Revelles2013}: 
\begin{equation}\label{eq:gamma_par_red}
\gamma(z)=\gamma_{0}+\gamma_{1}\frac{z}{1+z}
\end{equation}

This approach has been widely used to constrain the $f(R)$ gravity, pointing out that: the current data have not enough constraining power to discriminate between different modified gravity models and the $\Lambda$CDM model; 
the so-called growth index is the most efficient way to provide constraints on the model parameters;
and the {$\gamma$-parameterization} represents the best way to measure deviation from the standard model
\cite{Zhang2012, Revelles2013}. 

$\bullet$ {Dynamical variables:} Starting from Equations~\eqref{eq:mod_friedmann1} and
\eqref{eq:mod_friedmann2}, it is possible to reach a model-independent description
of the evolution of the matter density perturbations \cite{Amendola2007, Euclid2013, Tsujikawa2008}. 
Defining the following dimensionless variables:
\begin{align}
& x_1 \equiv-\frac{\dot{f}_{,R}}{Hf_{,R}}&\\
& x_2 \equiv-\frac{f(R)}{6H^2 f_{,R}}&\\
& x_3 \equiv\frac{R}{6H^2}&\\
& x_4 \equiv-\frac{\kappa^2\rho_\gamma}{3H^2 f_{,R}} \label{eq:x4}&
\end{align}
the density parameters and the effective equation of state of the system can be written as:
\begin{align}
&\Omega_\gamma\equiv x_4& \label{eq:dens1}\\
&\Omega_m\equiv\frac{\kappa^2\rho_m}{3H^2 f_{,R}}=1-x_1-x_2-x_3-x_4& \label{eq:dens2}\\
&\Omega_X\equiv x_1+x_2+x_3& \label{eq:dens3}\\
& w_{eff}\equiv-\frac{1}{3}(2x_3-1)
\end{align}

The background evolution is given by:
\begin{align}
& x'_1 = -1-x_3-3x_2+x_1^2-x_1x_3+x_4\label{eq:ev1}&\\
& x'_2 = \dfrac{x_1x_3}{m}-x_2(2x_3-4-x_1)\label{eq:ev2}&\\
& x'_3 = -\dfrac{x_1x_3}{m}-2x_3(x_3-2)\label{eq:ev3}&\\
& x'_4 = -2x_3x_4-x_1x_4\label{eq:ev4}&
\end{align}
where the prime indicates the derivative with respect to the scale factor $d/d\ln a$, and $m$ is a 
parameter related to the theory that indicates how much a specific model is close to the 
$\Lambda$CDM (or how much it deviates from it). This parameter is defined as:
\begin{align}
m \equiv \dfrac{d \ln f_{,R}}{d ln R}=\dfrac{R f_{,RR}}{f_{,R}}\label{eq:par1}
\end{align}
and by definition, one also finds the following identity:
\begin{align}
 r \equiv - \dfrac{d \ln f}{d ln R}=-\dfrac{R f_{,R}}{f}=\dfrac{x_3}{x_2}\label{eq:par2}
\end{align}

Equations~\eqref{eq:ev1} to \eqref{eq:par2} represent a closed system that can be integrated numerically, 
obtaining the evolution of the background densities perturbations. We just need to choose an $f(R)$-model 
to compute the two parameters $m$ and $r$ that we need to integrate the system. From the analysis of the critical 
points, we have several conditions that an $f(R)$ model has to respect to be cosmologically \mbox{viable~\cite{Amendola2007, Euclid2013}}. In~general, only models with $m\ge0$ and very close to the $\Lambda$CDM model ($f(R)=R-2\Lambda$)
are cosmologically viable. Thus, the equations for the matter density perturbations are~\cite{Tsujikawa2008, Hwang:2001qk}:
\begin{align}
\delta_{m}'' & +\left(x_{3}-\frac{1}{2}x_{1}\right)\delta_{m}'-\frac{3}{2}(1-x_{1}-x_{2}-x_{3})\delta_{m}\nonumber \\
 & =\frac{1}{2}\biggl[\left\{ \frac{k^{2}}{x_{5}^{2}}-6+3x_{1}^{2}-3x_{1}'-3x_{1}(x_{3}-1)\right\} \delta\tilde{f_{,R}}\nonumber \\
 & ~~~~+3(-2x_{1}+x_{3}-1)\delta\tilde{f_{,R}}'+3\delta\tilde{f_{,R}}''\biggr]\,\\
\delta\tilde{f_{,R}}'' & +(1-2x_{1}+x_{3})\delta\tilde{f_{,R}}'\nonumber \\
 & +\left[\frac{k^{2}}{x_{5}^{2}}-2x_{3}+\frac{2x_{3}}{m}-x_{1}(x_{3}+1)-x_{1}'+x_{1}^{2}\right]\delta\tilde{f_{,R}}\nonumber \\
 & ~~~~=(1-x_{1}-x_{2}-x_{3})\delta_{m}-x_{1}\delta_{m}'\,\end{align}
 where $\delta\tilde{f_{,R}}\equiv\delta f_{,R}/f_{,R}$ and $x_{5}\equiv aH$ satisfies
:
 \begin{eqnarray}
x_{5}'=(x_{3}-1)\, x_{5}\,.\label{x5eq}
\end{eqnarray}

The evolution at all scales of the growth of the matter density perturbation, $\delta_m$, could be obtained 
via numerical integration. However, a more straightforward approach is to assume that 
the growth rate, $f_g$, is a function of time, but not of scale \cite{peebles76,lahav91,polarski08,linder05,wang98},
and to use the {$\gamma$-parameterization} in Equations~\eqref{eq:gamma_par}~and~\eqref{eq:gamma_par_red}.

\begin{figure}[H]
\centering
\includegraphics[width=0.6\columnwidth]{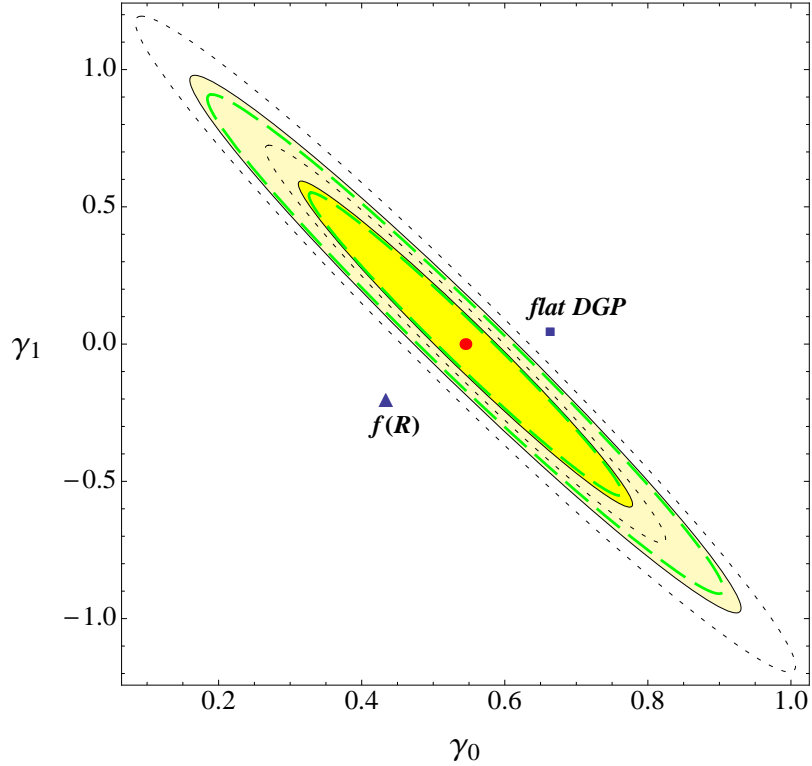} 
\caption{\label{fig:growth_contours}
The 1$\sigma$ and 2$\sigma$ marginalized contours 
for the parameters $\gamma_{0}$ and $\gamma_{1}$ in the 
$\gamma$-parameterization. Shown is the {reference case} (shaded yellow regions), 
with the {optimistic} error bars (green long-dashed ellipses) and the {pessimistic} ones 
(black dotted ellipses). Red circle represent the $\Lambda$CDM model ($\gamma=0.545$), while triangles 
represent the $f(R)$ model~\cite{Euclid2013}.\vspace{-6pt}} 
\end{figure}

The detection of possible deviation from $\Lambda$CDM model has been studied fitting
the $\gamma$-parameterization to the forthcoming Euclid dataset \cite{Euclid2013}.
Assuming the $\Lambda$CDM model as the reference model, 
it has been found that: the forthcoming data will be able to constrain the parameter 
$\gamma$ and $w$ with an accuracy of 4\% and 2\% if they do not depend on redshift 
(standard evolution) and to clearly distinguish the $\Lambda$CDM from 
alternative scenarios at more than $2\sigma$; see Figure \ref{fig:growth_contours}. 

\section{Testing Gravity Using the Cosmic Microwave Background Data}
\label{sec.6}

The CMB temperature anisotropies play a key rule 
to constrain the cosmological model. They are mostly generated at the 
last-scattering surface, and thus, they provide a way to probe the early 
Universe~\cite{Planck2015_13, Planck2015_20, Planck2015_27}. However, the lack of a 
full-consistent theoretical framework and the {fine tuning} problem related to 
the cosmological constant, $\Lambda$, have encouraged studying
wide classes of alternative DE or modified gravity models. The most
difficult challenge of modern cosmology is to use current and 
forthcoming datasets to discriminate between those alternative visions
of the Universe \cite{Euclid2013, LSST2009, Clifton2012, Joyce2015, Huterer2015}.
Many of those models mimic the $\Lambda$CDM 
at background evolution, but differ in the perturbations, and 
they could produce several effects on the CMB power spectrum.
As an example, different models could have different expansion 
histories that shift the location of the peaks \cite{Hu1996},
and could affect the gravitational potential at late time, changing the 
integrated Sachs--Wolfe (ISW) effect \cite{Sachs1967, Kofman1985}. Moreover, modification of the 
gravitational theory should be reflected in modifications of the 
lensing potential \cite{Acquaviva2006, Carbone2013}, the growth of the structures 
\cite{Peebles1984, Barrow1993, Kunz2004, Baldi2011} and also the amplitude of the primordial B-modes
\cite{Amendola2014, Raveri2014}.

Modified gravity models affect the evolution of the Universe at both
the background and the perturbation \cite{abebe2012,abebe2013,Carloni2008,Ananda2009}. There are mainly two different approaches used
to constrain alternative models of gravity: the parametrized modified gravity approach and
the effective field theory.

The parametrized modified gravity approach  constructs functions probing the geometry of space-time 
and the growth of perturbations. Starting from a spatially-flat FLRW metric, the perturbed line 
element in the Newtonian gauge is given by:
\begin{equation}\label{eq:metric}
ds^2 = a(\tau)^2[-(1+2\Psi)d\tau^2+(1-2\Phi)dx^2] \,
\end{equation}
where $\tau$ is the conformal time. $\Psi$ and $\Phi$ are the two scalar gravitational 
potentials that are a function of the scale $\Psi=\Psi(k,a)$ and $\Phi=\Phi(k,a)$
and are related to the energy-momentum tensor, $T_{\mu\nu}$, through Einstein's equations.
Those potentials can be modeled to fix all degrees of freedom in order to match observational data.
Moreover, the difference of those two potentials $\Phi-\Psi$, which is the anisotropic stress,
has to be equal to zero in GR. Any departure from this value would indicate a departure from GR~\cite{Mukhanov1992, Saltas2014}. One of the most common parameterization is 
implemented in the {Modification of Growth with Code for Anisotropies in the Microwave Background} (the 
code is publicly available at {http://www.sfu.ca/~aha25/MGCAMB.html}) (MGCAMB) \cite{Lewis:1999bs, Zhao:2008bn}, 
and it considers the following:
\begin{align}
\label{eq:parametrization-Poisson}
& k^2\Psi=-4 \pi G a^2\mu(k,a)\rho\Delta \, \\
\label{eq:parametrization-anisotropy}
& \frac{\Phi}{\Psi}=\gamma(k,a) \,
\end{align}
to parameterize the Poisson and the anisotropic stress equations using two scale and time-dependent 
functions $\mu(k,a)$ and $\gamma(k,a)$. Those two functions are determined by choosing the particular
model. For~example, some classes of $f(R)$ models lead to \cite{Bertschinger2008}:
\begin{eqnarray}
&& \mu(k,a)=\frac{1+\beta_1\lambda_1^2\,k^2a^s}{1+\lambda_1^2\,k^2a^s} \, \\
\label{BZ}
&& \gamma(k,a)=\frac{1+\beta_2\lambda_2^2\,k^2a^s}{1+\lambda_2^2\,k^2a^s} \
\end{eqnarray}
where the parameters $\beta_i$ are dimensionless and represent the couplings,
and the parameters $\lambda_i$ are lengths. Those functional forms have been used 
in early studies to forecast errors of the parameters for the 
Dark Energy Survey (www.darkenergysurvey.org/) (DES, \cite{Abbott2005}) and 
Large Synoptic Survey Telescope ({http://www.lsst.org/lsst/}) (LSST, \cite{Ivezic}) 
surveys \cite{Zhao:2008bn}.
The main limitation of this approach is the use of the quasi-static regime assuming
that the scales at which one is interested are still linear and smaller 
than the horizon. Furthermore, the time derivatives are neglected.

{Effective field theory (EFT)} \cite{Cheung2008, Gubitosi2013} 
describes the whole range of the scalar field theories. The 
Lagrangian preserves the isotropy and homogeneity of the Universe at the background, 
providing a more general approach without assuming the quasi-static regime, but
adding much more free parameters that must be constrained. The $f(R)$-models are a 
sub-class of those theories. The action of EFT reads:
\begin{eqnarray} \label{eq:EFTLag}
 S &=& \int d^4x\sqrt{-g}\left\{\frac{m_0^2}{2}\left[1+\Omega(\tau)\right]R
 + \Lambda(\tau) - a^2c(\tau) \delta g^{00} \right. \nonumber \\ 
 &\quad&\left. +\ \frac{M_2^4 (\tau)}{2} \left(a^2\delta g^{00}\right)^2 
 - \bar{M}_1^3 (\tau){2}a^2 \delta g^{00} \delta {K}^\mu_\mu \right.
 \nonumber \\ 
 &\quad& \left. -\ \frac{\bar{M}_2^2 (\tau)}{2} \left(\delta K^\mu_\mu\right)^2
 - \frac{\bar{M}_3^2 (\tau)}{2} \delta K^\mu_\nu \delta K^\nu_\mu
 +\frac{a^2\hat{M}^2(\tau)}{2}\delta g^{00}\delta R^{(3)} \right. \nonumber \\ 
 &\quad&\left. +\ m_2^2 (\tau) \left(g^{\mu \nu} + n^\mu n^\nu\right)
 \partial_\mu \left(a^2g^{00}\right) \partial_\nu\left(a^2 g^{00}\right)
 \vphantom{\frac{m_0^2}{2}}\right\} \nonumber \\
 &\quad& +\ S_{\rm{m}} \big[\chi_i ,g_{\mu \nu}\big]
\end{eqnarray} 
where $\delta R^{(3)}$ is its spatial perturbation,
${K}^\mu_\nu$ is the extrinsic curvature, $m_0$ is the bare (reduced)
Planck mass and $S_{\rm{m}}$ is the matter part of the action, including all fluid
components, such as baryons, cold DM, radiation,
and neutrinos, but it does not include DE. The nine
time-dependent functions~\cite{Bloomfield2013}
$\{ \Omega,c,\Lambda,\bar{M}_1^3,\bar{M}_2^4,\bar{M}_3^2,M_2^4,\hat{M}^2,
 m_2^2 \}$, have to be fixed to specify the theory. The EFT 
has been implemented in the publicly-available {\tt EFTCAMB}
 code ({http://www.lorentz.leidenuniv.nl/\~hu/codes/}, Version 1.1, October 2014.) \cite{EFTCAMB1, EFTCAMB2}, 
that numerically solves both the background and perturbation equations. The code 
has been used to constrain massive neutrinos in $f(R)$ gravity \cite{EFTCAMB1}.

Planck collaboration has used in both approaches to fit $f(R)$ models to 
investigate their effect combining early time datasets, such as CMB, 
and cosmological datasets at much lower redshift \cite{Planck2015_14}. In general, $f(R)$ models
require setting the initial condition:
\begin{equation}
\lim_{R \to \infty} \frac{f(R)}{R} = 0
\end{equation}
and the boundary condition:
\begin{equation}
\label{eqn:b0}
B(z) = \frac{f_{RR}}{1+f_R}\frac{H \dot{R}}{\dot{H}-H^2} 
\end{equation}

\begin{figure}[H]
\centering
\includegraphics[width=.6\textwidth]{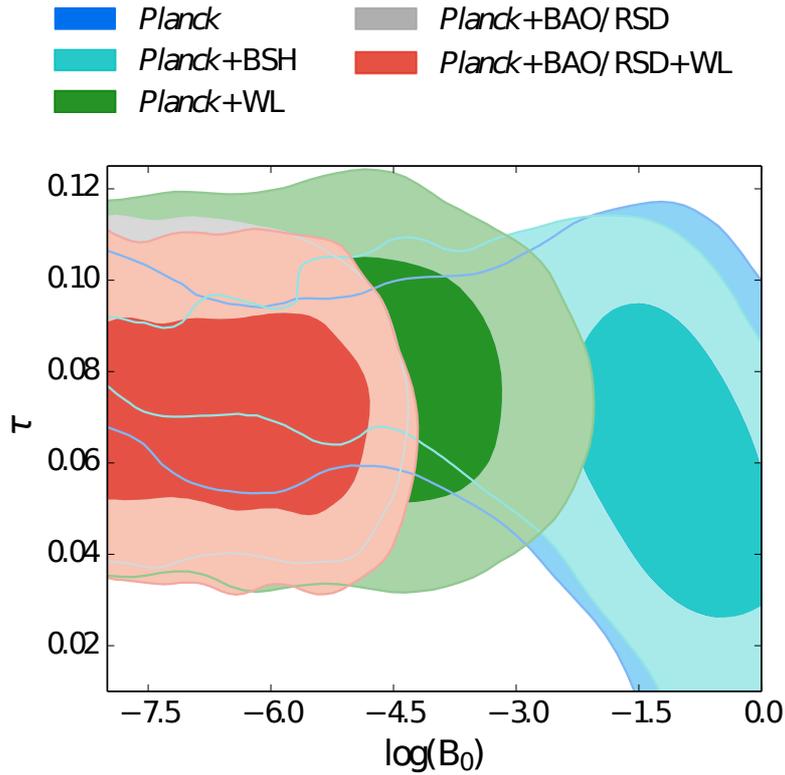}
\caption{The $68\%$ and $95\%$ contour plots for the two parameters,
$\{\rm{Log}_{10}(B_0),\tau\}$.
There is a degeneracy between the two parameters for {Planck} temperature power spectrum (TT)
 + BSH (the combination of BAO, SNIa, and $H_0$ datasets). Adding
lensing will break the degeneracy between the two. Here, {Planck}
indicates {Planck} TT.}
\label{fig:logB0-tau}

\end{figure}

High values of $B_0$, its present value, change the ISW effect and the CMB lensing
\mbox{potential~\cite{Song2007,Schmidt2008, Bertschinger2008},
Marchini2013}. Their analysis has shown the presence of a degeneracy
between the optical depth $\tau$ and $B_0$ that can be broken adding 
other probes on the structure formation, such as weak lensing, CMB lensing
or redshift-space distortions (see Figure \ref{fig:logB0-tau}).
The same study carried out using MGCAMB code led to the same results
within the uncertainties, improving the previous constraint on $B_0$
\cite{Song2007a}.

Nevertheless, both EFTCAMB and MGCAMB assume a specific background cosmology. MGCAMB code
assumes for the background evolution the $\Lambda$CDM or $w$CDM, while EFTCAMB assumes a fixed background
evolution, and the CMB power spectrum is calculated for an $f(R)$ model re-constructed by the fixed
expansion history. A more recent code, named 
$f(R)$ Code for Anisotropies in the Microwave Background (FRCAMB)
 \cite{Lixin2015}, allows reconstructing the CMB power 
spectrum for any $f(R)$ model by specifying the first and second derivatives of the Lagrangian. Although 
the results obtained are consistent with the one from EFTCAMB and MGCAMB, the code has the advantage
of being designed for any $f(R)$ model at both the background and perturbation level.

\section{Discussion and Future Perspectives}
\label{sec.7}

In this review, we have outlined the possibilities offered by the current and the forthcoming 
astrophysical and cosmological dataset to constrain or rule out the $f(R)$-models. 
It is mandatory to search for the answer to one of the fundamental questions in modern cosmology:
is GR the effective theory of gravitation? This question comes from the evidence that GR 
is not enough to fully explain the cosmological evolution of the Universe and the emergence
of the clustered structures, and it needs the addition of two unknown
components. The difficulties in identifying the DM and DE components at a fundamental level 
have opened the possibility that the answer resides in modifying the theory of gravity.
Nevertheless, having an alternative demands testing
it study the physical phenomena at much smaller scales than the cosmological ones.
To this effect, much analysis has been carried out to ensure that modified gravity 
could recover the tight constraints at the Solar System scale. Models that survive these probes have been
used to test galactic and extragalactic scales. Clusters of galaxies provide 
an excellent laboratory to probe the different features of the different modified 
gravity models. The variety of the phenomenology related to the cluster physics allows one to use
SZ, X-ray and WL observations in order to constrain $f(R)$-models. 
Many efforts have been also devoted to quantitatively describing the physical effects of alternative
cosmological scenarios. Those efforts have required developing new algorithms for hydrodynamical N-body simulations, to study the degeneracy between baryonic physics and modifications of gravity and also 
new Boltzmann codes to predict the CMB power spectrum and to take advantage of the {Planck}
data to constrain modified gravity models. 

Nevertheless, a self-consistent picture has not been reached yet. The current datasets
do not have enough statistical power to discriminate between modified gravity models and standard cosmology,
and~when the needed precision is present, several degeneracies have been found. 
Next generation experiments, such as the
LSST \cite{Ivezic} and  the Wide-Field Infrared Survey Telescope (WFIRST) 
 ({http://wfirst.gsfc.nasa.gov/) \cite{wfirst}, will undertake large imaging surveys of the sky,
while other surveys, such as Euclid \cite{Laureijs2011} and the Big-Baryon Oscillation Spectroscopic Survey (BigBOSS)
 \cite{bigboss}, will make
spectroscopic measurements of galaxies. The combination of all forthcoming datasets will
hopefully put strong constraints on the evolution of the Universe and the 
emergence of the LSS, allowing us to discriminate between the {concordance} $\Lambda$CDM model
and its alternatives.



\acknowledgments{Acknowledgments}
Mariafelicia De Laurentis and Salvatore Capozziello acknowledge the Instituto Nazionale Fisica Nucleare (INFN)
 Sez. di Napoli (Iniziative Specifiche
QGSKY
 and TEONGRAV) 
for financial support.


\conflictofinterests{Conflicts of Interest}
The authors declare no conflict of interest. 


\end{document}